\documentclass[graybox]{svmult}

\usepackage{helvet}         
\usepackage{courier}        
\usepackage{type1cm}        
\usepackage{makeidx}         
\usepackage{graphicx}        
\usepackage{multicol}        
\usepackage[bottom]{footmisc}

\usepackage{hyperref,graphicx}
\usepackage{dcolumn}
\usepackage{bm}
\usepackage{bbm}
\usepackage{tikz}
\usepackage{physics}
\usepackage{enumitem}
\usepackage{amssymb,amsfonts}
\usepackage{epsfig}
\usepackage{epstopdf}
\usepackage[all]{xy}
\usepackage{dcolumn}
\usepackage{hyperref}
\usepackage{url}
\usepackage{dsfont}
\usepackage{slashed}
\usepackage{mathrsfs}
\usepackage{soul}
\usepackage{float}
\DeclareMathAlphabet{\mathpzc}{OT1}{pzc}{m}{it}

\newcommand{\be}{\begin{equation}}
\newcommand{\ee}{\end{equation}}
\newcommand{\bea}{\begin{eqnarray}}

\newcommand{\eea}{\end{eqnarray}}

\def\inbar{\,\vrule height1.5ex width.4pt depth0pt}
\def\IR{\relax{\rm I\kern-.18em R}}
\def\IC{\relax\hbox{$\inbar\kern-.3em{\rm C}$}}

\begin{document}

\title*{de Sitter Relativity Group}
\author{Hamed Pejhan}
\institute{Hamed Pejhan \at Institute of Mathematics and Informatics, Bulgarian Academy of Sciences, Acad. G. Bonchev Str. Bl. 8, 1113, Sofia, Bulgaria, \email{pejhan@math.bas.bg} \at This work is supported by the Bulgarian Ministry of Education and Science, Scientific Programme ``Enhancing the Research Capacity in Mathematical Sciences (PIKOM)", No. DO1-67/05.05.2022.}
\maketitle

\abstract{Building upon the comprehensive framework outlined in the author's recent collaborative book \cite{1}, this manuscript delivers a brief overview of the $1+4$-dimensional de Sitter (dS$_4$) group, its accompanying Lie algebra, and the corresponding (co-)adjoint orbits, with the latter assuming significance as potential classical elementary systems within the framework of dS$_4$.}

\section{dS$_4$ manifold and its relativity group} \label{sec:1}

The dS$_4$ spacetime exhibits a topological structure represented as ${\mathbb{R}}^1 \times {\mathbb{S}}^3$, where ${\mathbb{R}}^1$ corresponds to a timelike direction. One can easily picture it as a single-sheeted hyperboloid that is embedded within a $1+4$-dimensional Minkowski spacetime $\mathbb{R}^5$:
\begin{align}\label{MR}
{M}_R = \left\{x \in\mathbb{R}^5 \;;\; (x)^2 = \eta^{}_{\alpha\beta}x^\alpha x^\beta = -R^2 \right\}\,,
\end{align}
where $x^\alpha$ ($\alpha,\beta = 0,1,2,3,4$) represents the respective Cartesian coordinates and $\eta^{}_{\alpha\beta} = \mbox{diag}(1,-1,-1,-1,-1)$ denotes the ambient Minkowski metric.

The relativity group of dS$_4$ spacetime, analogous to the Lorentz group of the ambient Minkowski spacetime $\mathbb{R}^5$, is denoted as SO$_0(1,4)$. It constitutes a ten-parameter group encompassing linear transformations in $\mathbb{R}^5$ that maintain the invariance of the quadratic form $(x)^2 = \eta^{}_{\alpha\beta}x^\alpha x^\beta$, possess determinant $1$, and do not invert the direction of the ``time'' variable $x^0$.

The \emph{universal covering group} of SO$_0(1,4)$ is the symplectic Sp$(2,2)$ group. This term ``universal covering'' indicates that Sp$(2,2)$ serves as the covering group of SO$_0(1,4)$ and is simply connected. The Sp$(2,2)$ group becomes prominent when addressing systems involving half-integer spins. It is analogous to how SU$(2)$ replaces SO$(3)$ or SL$(2,\mathbb{C})$ replaces SO$_0(1,3)$. 
  
The Sp$(2,2)$ group can be appropriately characterized as the set of all $2 \times 2$-matrices ${g}$, with \emph{(real) quaternionic components}\footnote{In this manuscript, we exclusively focus on the $2\times 2$-matrix representation of quaternions. Within this representation, the quaternionic basis is defined as follows: $\big\{ \textbf{1} \equiv \mathbbm{1}_2, {\textbf{e}}^{}_k \equiv (-1)^{k+1} \mathrm{i} \sigma^{}_k \;;\; k=1,2,3 \big\}$, where $\sigma^{}_k$ represents the Pauli matrices.}, that satisfy both the unimodular and pseudo-unitary conditions:
\begin{eqnarray}\label{sp22somq}
\mathrm{Sp}(2,2) = \left\{{g} =
\begin{pmatrix}
\textbf{a} & \textbf{b}\\
\textbf{c} & \textbf{d}
\end{pmatrix}
\; ; \;\; \textbf{a},\textbf{b},\textbf{c},\textbf{d}\in\mathbb{H}, \; \mbox{det}({g}) = 1,\;{g}^\dagger \gamma^0 {g} = \gamma^0 \right\}\,,
\end{eqnarray}
where ${g}^\dagger = {{g}^{\scriptscriptstyle\bigstar}}^{\texttt{t}}$, such that ${g}^{\scriptscriptstyle\bigstar}$ refers to the quaternionic conjugate of ${g}$ and ${{g}}^{\texttt{t}}$ to the transpose of ${g}$, and $\gamma^0 = \begin{pmatrix} \textbf{1} & \textbf{0} \\ \textbf{0} & \textbf{-1} \end{pmatrix}$ in which the components $\textbf{1}$ and $\textbf{0}$ are, respectively, the unit and zero $2\times 2$ matrices.

The matrix $\gamma^0$ is an element of the Clifford algebra, as established by:
\begin{equation}\label{clifford}
\gamma^\alpha \gamma^\beta + \gamma^\beta \gamma^\gamma = 2\eta^{\alpha\beta} \mathbbm{1}_4\,, \quad {\gamma^\alpha}^\dagger = \gamma^0 \gamma^\alpha \gamma^0\,,
\end{equation}
where, with quaternionic components, the other four matrices are defined by:
\begin{equation}\label{gamexp}
\gamma^k = (-1)^{k+1} \begin{pmatrix} \textbf{0} & \mathrm{i} \sigma_k \\ \mathrm{i} \sigma_k & \textbf{0} \end{pmatrix} \equiv 
\begin{pmatrix} \textbf{0} & {\textbf{e}}^{}_k \\ {\textbf{e}}^{}_k & \textbf{0} \end{pmatrix}\,, \quad \gamma^4 = \begin{pmatrix} \textbf{0} & \textbf{1}\\ -\textbf{1} & \textbf{0} \end{pmatrix}\,,
\end{equation}
with $\sigma_k$ ($k=1,2,3$) denoting the Pauli matrices. In the given expressions, $\mathbbm{1}_4$ represents the $4\times 4$-unit matrix. Among the $\gamma^\alpha$ matrices, it is important to note that $\gamma^0$ is the sole member belonging to Sp$(2,2)$.

A common way to represent the associated Lie algebra is by considering the linear span generated by the ten Killing vectors:
\begin{eqnarray}\label{Killing dS4}
K_{\alpha\beta} = x_\alpha \partial_\beta - x_\beta \partial_\alpha\,, \quad K_{\alpha\beta} = - K_{\beta\alpha}\,.
\end{eqnarray}
They obey the following commutation relations:
\begin{eqnarray}\label{algebra}
\left[K_{\alpha\beta},K_{\rho\delta}\right] = - \left( \eta^{}_{\alpha\rho} {K_{\beta\delta}} + \eta^{}_{\beta\delta} {K_{\alpha\rho}} - \eta^{}_{\alpha\delta} {K_{\beta\rho}} - \eta^{}_{\beta\rho} {K_{\alpha\delta}} \right)\,.
\end{eqnarray}

\section{A discrete symmetry and energy ambiguity} \label{sec:2}
Let us associate with any $x \in \mathbb{R}^5$ the matrix $\slashed{x}$ as given below:
\begin{eqnarray}\label{slashx}
\slashed{x} = x^\alpha \gamma_\alpha =
\begin{pmatrix}
\textbf{1} x^0 & - \textbf{x} \\
{\textbf{x}}^{\scriptscriptstyle\bigstar} & - \textbf{1} x^0
\end{pmatrix}\,,
\end{eqnarray}
where, in the scalar-vector representation of a quaternion, $\textbf{x} = (x^4, \vec{x}) \in \mathbb{H}$ (equivalently, in the Euclidean metric notations, $\textbf{x} = x^4\textbf{1} + x^1{\textbf{e}}^{}_1 + x^2{\textbf{e}}^{}_2 + x^3{\textbf{e}}^{}_3$). Conversely, a matrix in the form of $\slashed{x}$ uniquely maps to a point $x \in \mathbb{R}^5$, with components determined as $x^\alpha = \frac{1}{4} \mbox{tr}\left(\gamma^\alpha \slashed{x}\right)$. As a result, this correspondence establishes a one-to-one mapping between $\mathbb{R}^5$ and the collection of $4\times 4$-matrices $\slashed{x}$. Then, the action of $g\in \text{Sp}(2,2)$ on $x\in\mathbb{R}^5$ is defined by:
\begin{eqnarray}\label{xxxxxxxxx}
\slashed{x}^\prime = {g} \slashed{x} {g}^{-1} =
\begin{pmatrix}
\textbf{1} x^{\prime 0} & -\textbf{x}^{\prime}\\
{\textbf{x}}^{\prime{\scriptscriptstyle\bigstar}} & - \textbf{1} x^{\prime 0}
\end{pmatrix}\,.
\end{eqnarray}

An intriguing aspect worth emphasizing here is that within the context of Sp$(2,2)$, the group action of $\gamma^0$ aligns with a discrete symmetry:
\begin{equation}\label{actgam0}
\slashed{x} \;\mapsto\; \slashed{x}^\prime = \gamma^0 \slashed{x} \left(\gamma^0\right)^{-1} = \begin{pmatrix}
\textbf{1} x^0 & \textbf{x} \\
-{\textbf{x}}^{\scriptscriptstyle\bigstar} & - \textbf{1} x^0
\end{pmatrix}\,.
\end{equation}
This transformation mirrors any point $x=(x^0,\textbf{x}) \in{M}_R$ with respect to the $x^0$-axis, such that $x=(x^0, \mathbf{x})$ is mapped to its mirror image $x^{\prime}=(x^0,-\textbf{x})$.

Under this discrete symmetry, the dS$_4$ infinitesimal generators $K_{a0}$, with $a= 1, 2, 3, 4$, undergo transformations that result in their counterparts, potentially with differing signs for the associated conserved charges, depending on the sign of $\textbf{x}$. This situation, for instance, suggests that the movement induced by the generator $K_{40}$, which contracts to the Poincar\'{e} energy operator, in the temporal dimension (whether it propels us forward or backward with respect to the increase or decrease in $x^0$) is contingent upon the sign of $\textbf{x}$, resulting in an inability to precisely determine its direction. This fact underscores the fundamental limitation that there is no preserved positive energy in dS$_4$, and more broadly, in dS spacetime.

\section{Space-time-Lorentz decomposition}

Any element ${g}\in\mathrm{Sp}(2,2)$, in consideration of the group involution $\mathfrak{i}({g}) : {g} \;\mapsto\; \gamma^0\gamma^4 {g}^\dagger \gamma^0\gamma^4$, can be decomposed in a nonunique way into:
\begin{align}
{g} =
\begin{pmatrix}
\textbf{a} & \textbf{b} \\
\textbf{c} & \textbf{d}
\end{pmatrix} = 
\underbrace{\begin{pmatrix}
{\textbf{w}} & \textbf{0} \\
\textbf{0} & {\textbf{w}}^{\scriptscriptstyle\bigstar}
\end{pmatrix}}_{\equiv\; T_{s.t.}(\textbf{w})}
\underbrace{\begin{pmatrix}
\textbf{1}\cosh \frac{\psi}{2} & \textbf{1}\sinh\frac{\psi}{2}\\
\textbf{1}\sinh\frac{\psi}{2} & \textbf{1}\cosh \frac{\psi}{2}
\end{pmatrix}}_{\equiv\; T_{t.t.}({\psi})} 
\underbrace{\begin{pmatrix}
\textbf{v} & \textbf{0} \\
\textbf{0} & \textbf{v}
\end{pmatrix}}_{\equiv\; T_{s.r.}(\textbf{v})}
\underbrace{\begin{pmatrix}
\textbf{1}\cosh \frac{\varphi}{2} & \vec{\textbf{u}}\sinh \frac{\varphi}{2}\\
- \vec{\textbf{u}}\sinh\frac{\varphi}{2} & \textbf{1}\cosh\frac{\varphi}{2}
\end{pmatrix}}_{\equiv\; T_{b.t.}(\varphi,\vec{\textbf{u}})}\,,\nonumber
\end{align}
where $\psi,\varphi\in\mathbb{R}$, ${\textbf{w}} \equiv \big( \cos\frac{\theta}{2}, \sin\frac{\theta}{2} \vec{w} \big)$, $\textbf{v} \equiv \big(\cos\frac{\vartheta}{2}, \sin\frac{\vartheta}{2} \vec{v}\big)$, and the pure vector quaternion $\vec{\textbf{u}} \equiv \big( 0, \vec{u} \big)$, with $0\leqslant \theta,\vartheta < 2\pi$ and $\vec{w}, \vec{v}, \vec{u}\in\mathbb{S}^2$ (and hence, $\textbf{w}, \textbf{v} \in\mathbb{S}^3$). In this group decomposition:
\begin{enumerate}
    \item{The transformations $T_L(\textbf{v},\varphi,\vec{\textbf{u}}) = T_{s.r.}(\textbf{v}) \, T_{b.t.}(\varphi,\vec{\textbf{u}})$ leave invariant the point $x^{}_\odot = (0,0,0,0,R)$, selected as the origin of the dS$_4$ hyperboloid $M_R$. The tangent space at this point, i.e., the hyperplane $\{x\in\mathbb{R}^5 \; ;\; x^4 = R \}$, is a $1+3$-dimensional Minkowski spacetime with pseudo-metric $\mathrm{d} x_0^2 - \mathrm{d} x_1^2 - \mathrm{d} x_2^2 - \mathrm{d} x_3^2$. It serves as the limit where dS$_4$ spacetime contracts to zero curvature. Thus, the subgroup $T_L(\textbf{v},\varphi,\vec{\textbf{u}})$, isomorphic to $\mathrm{SO}_0(1,3)$, which is the stabilizer of $x^{}_\odot$, can be interpreted as the Lorentz group for the tangent space. This clarifies the understanding of the transformations produced by $T_{s.r.}(\textbf{v})$ and $T_{b.t.}(\varphi,\vec{\textbf{u}})$. They correspond to ``space rotations" and ``boost transformations", respectively. Then, it is assumed that the parameters $\textbf{v}$, $\vec{\textbf{u}}$, and $\varphi$ respectively represent space rotation, boost velocity direction, and rapidity.}
    \item{The transformations $T_{\text{dS}_4}(\textbf{w},\psi) = T_{s.t.}(\textbf{w})\, T_{t.t.} ({\psi})$ map the origin $x^{}_\odot$ to every point within ${M}_R$, effectively covering the entire dS$_4$ manifold. In this sense, the transformations $T_{s.t.}(\textbf{w})$ and $T_{t.t.} ({\psi})$ are regarded as the ``space translations" and ``time translations", respectively.}
\end{enumerate}

Given the space-time-Lorentz decomposition of the group, the associated infinitesimal generators can be expressed as:
\begin{eqnarray}
\label{Xk} {X}_k &=& \frac{\mathrm{d}}{\mathrm{d}\theta}
\begin{pmatrix}
{\textbf{w}}^{}_k & \textbf{0} \\
\textbf{0} & {\textbf{w}}^{\scriptscriptstyle\bigstar}_k
\end{pmatrix}\bigg|_{\theta=0} = \frac{1}{2}
\begin{pmatrix}
{\textbf{e}}^{}_k & \textbf{0} \\
\textbf{0} & - {\textbf{e}}^{}_k
\end{pmatrix}\,, \quad {\textbf{w}}^{}_k \equiv \left( \cos\textstyle\frac{\theta}{2}, \sin\frac{\theta}{2}\, {\textbf{e}}^{}_k \right)\,, \\
\label{X0} {X}_0 &=& \frac{\mathrm{d}}{\mathrm{d}\psi}
\begin{pmatrix}
\textbf{1}\cosh \frac{\psi}{2} & \textbf{1}\sinh\frac{\psi}{2}\\
\textbf{1}\sinh\frac{\psi}{2} & \textbf{1}\cosh \frac{\psi}{2}
\end{pmatrix}\bigg|_{\psi=0} = \frac{1}{2}
\begin{pmatrix}
\textbf{0} & \textbf{1} \\
\textbf{1} & \textbf{0}
\end{pmatrix}\,, \\
\label{Yk} {Y}_k &=& \frac{\mathrm{d}}{\mathrm{d}\vartheta}
\begin{pmatrix}
{\textbf{v}}^{}_k & \textbf{0} \\
\textbf{0} & {\textbf{v}}^{}_k
\end{pmatrix}\bigg|_{\vartheta=0} = \frac{1}{2}
\begin{pmatrix}
{\textbf{e}}^{}_k & \textbf{0} \\
\textbf{0} & {\textbf{e}}^{}_k
\end{pmatrix}\,, \quad {\textbf{v}}^{}_k \equiv \left( \cos\textstyle\frac{\vartheta}{2}, \sin\frac{\vartheta}{2}\, {\textbf{e}}^{}_k \right)\,,\\
\label{Zk} {Z}_k &=& \frac{\mathrm{d}}{\mathrm{d}\varphi}
\begin{pmatrix}
\textbf{1}\cosh \frac{\varphi}{2} & \vec{\textbf{u}}^{}_k\sinh \frac{\varphi}{2}\\
- \vec{\textbf{u}}^{}_k\sinh\frac{\varphi}{2} & \textbf{1}\cosh\frac{\varphi}{2}
\end{pmatrix}\bigg|_{\varphi=0} = \frac{1}{2}
\begin{pmatrix}
\textbf{0} & {\textbf{e}}^{}_k \\
- {\textbf{e}}^{}_k & \textbf{0}
\end{pmatrix}\,, \quad \vec{\textbf{u}}^{}_k \equiv \big( 0, {\textbf{e}}^{}_k \big)\,.\quad\quad
\end{eqnarray}
Let $K_{4k} \equiv {X}_k$, $K_{04} \equiv {X}_0$, $K_{ki} \equiv {{\cal{E}}_{ki}}^{j} \,{Y}_j$, $K_{0k} \equiv {Z}_k$, where $i,j,k = 1,2,3$ and ${{\cal{E}}_{ij}}^{k}$ is the three-dimensional totally antisymmetric Levi-Civita symbol. It is then straightforward to confirm that these defined infinitesimal generators satisfy the dS$_4$ Lie algebra $\mathfrak{sp}(2,2)$ as indicated in Eq. \eqref{algebra}.

\section{$\text{d}$S$_4$ scalar elementary systems on the classical level}
The dS$_4$ Lie algebra $\mathfrak{sp}(2,2)$ in the introduced quaternionic notations can be represented as the linear span of the aforementioned infinitesimal generators:
\begin{align}\label{algebra dS4}
\mathfrak{sp}(2,2) =&\; \Bigg\{ 2a^k {X}_k + 2j^k {Y}_k + 2d^0 {X}_0 + 2d^k {Z}_k =
\begin{pmatrix} (a^k+j^k) \textbf{e}^{}_k & d^0 \textbf{1} + d^k \textbf{e}^{}_k \\ d^0 \textbf{1} - d^k \textbf{e}^{}_k & (-a^k+j^k) \textbf{e}^{}_k \end{pmatrix} \Bigg\}\,, \nonumber
\end{align}
where $a^k, j^k, d^0, d^k \in \mathbb{R}$ ($k=1,2,3$). Then, $\mathfrak{sp}(2,2)$ is in one-to-one correspondence with $\mathbb{R}^{10}$; $\mathfrak{sp}(2,2)\sim \mathbb{R}^{10}$.

On the classical level, understanding dS$_4$ elementary systems can be achieved through the conventional phase-space approach, a well-established method closely associated with the concept of Sp$(2,2)$ co-adjoint action, as detailed in Refs. \cite{Kirillov, Kirillov1976}. These elementary systems are effectively described in terms of orbits resulting from this action. These orbits, symplectic manifolds in their own right, possess a natural Sp$(2,2)$-invariant (Liouville) measure and are Sp$(2,2)$-homogeneous spaces. Specifically, they are homeomorphic to even-dimensional group cosets of the form $\mathrm{Sp}(2,2)/{\cal{S}}$, where ${\cal{S}}$ is the stabilizer subgroup associated with some orbit point. Here, it is essential to note that Sp$(2,2)$ being a simple group, its adjoint action on the Lie algebra $\mathfrak{sp}(2,2)$, that is:
\begin{eqnarray}\label{Ad_g dS4}
{g} \in \mathrm{Sp}(2,2), \; {X} \in \mathfrak{sp}(2,2) \; ; \; \mbox{Ad}_{{g}}({X}) = {g} {X} {g}^{-1}\,,
\end{eqnarray}
is equivalent to its co-adjoint action on the dual of $\mathfrak{sp}(2,2)$ \cite{Kirillov, Kirillov1976}.

The family of (co-)adjoint orbits associated with dS$_4$ scalar ``massive" elementary systems is linked to the transformation of the element $2 \kappa {X}_0 = \kappa\begin{pmatrix} \textbf{0} & \textbf{1} \\ \textbf{1} & \textbf{0} \end{pmatrix}$, where $0<\kappa<\infty$, through the (co-)adjoint action (\ref{Ad_g dS4}). In this context, the subgroup responsible for stabilizing the element $2 \kappa {X}_0$ comprises the space-rotations and time-translations subgroups, which stem from the space-time-Lorentz decomposition of Sp$(2,2)$. Consequently, this family can be described as the group coset $O(2 \kappa {X}_0) \sim \mathrm{Sp}(2,2)/\big(\mathrm{SU}(2) \times \mathrm{SO}_0(1,1)\big)$.

In the context of the space-time-Lorentz decomposition of Sp$(2,2)$, this description can be explicitly realized by applying the space-translations and Lorentz-boosts subgroups to transport the element $2 \kappa {X}_0$ under the effective (co-)adjoint action (\ref{Ad_g dS4}):
\begin{eqnarray}\label{zzzzz}
\mbox{Ad}_{{g}}(2 \kappa {X}_0) &=& T^{}_{s.t.}(\textbf{w})\, T^{}_{b.t.}(\varphi,\vec{\textbf{u}}) \; \Big( 2 \kappa {X}_0 \Big)\; T^{-1}_{b.t.}(\varphi,\vec{\textbf{u}})\, T^{-1}_{s.t.}(\textbf{w}) \nonumber\\
&=&
\begin{pmatrix}
\vec{\textbf{p}} & p_0 {\textbf{w}}^2 \\
p_0 {\textbf{w}}^{2{\scriptscriptstyle\bigstar}} & -{\textbf{w}}^{2{\scriptscriptstyle\bigstar}} \vec{\textbf{p}} {\textbf{w}}^2
\end{pmatrix} \equiv \begin{pmatrix}
\vec{\textbf{p}} & p_0 {\textbf{z}} \\
p_0 {\textbf{z}}^{{\scriptscriptstyle\bigstar}} & -{\textbf{z}}^{{\scriptscriptstyle\bigstar}} \vec{\textbf{p}} {\textbf{z}}
\end{pmatrix}\equiv {X}({\textbf{z}},\vec{\textbf{p}}) \,,
\end{eqnarray}
where $\vec{\textbf{p}}\equiv \kappa {\textbf{w}} \vec{\textbf{u}} {\textbf{w}}^{\scriptscriptstyle\bigstar} \sinh{\varphi}$ represents a pure vector quaternion and $p_0 = \kappa\cosh\varphi = \left(\kappa^2 + \big|\vec{\textbf{p}}\big|^2\right)^{1/2}$. This parameterization elucidates the inherent topological structure, denoted as ${\mathbb{S}}^3 \times {\mathbb{R}}^3 = \big\{ {X}({\textbf{z}},\vec{\textbf{p}}) \; ; \; {\textbf{z}} \in \mathrm{SU}(2) \sim {\mathbb{S}}^3,\; \vec{\textbf{p}} \sim {\mathbb{R}}^3 \big\}$, of the ($6$-dimensional) (co-)adjoint orbits $O(2 \kappa {X}_0)$. Note that the invariant measure associated with the coordinates $({\textbf{z}},\vec{\textbf{p}})$ can be expressed as:
\begin{eqnarray}
\mathrm{d}\mu ({\textbf{z}},\vec{\textbf{p}}) = \mathrm{d}\mu({\textbf{z}}) \; \mathrm{d}^3 \vec{\textbf{p}} \,,
\end{eqnarray}
where $\mathrm{d}\mu({\textbf{z}})$ represents the $O(4)$-invariant measure on $\mathbb{S}^3$.

Taking into account $a^k, j^k, d^0,$ and $d^k$ ($k=1,2,3$) as cartesian coordinates on the dual of $\mathfrak{sp}(2,2)$ (again, the Lie algebra of a simple Lie group, and in general, a semi-simple Lie group, exhibits an isomorphism with its dual):
\begin{eqnarray}\label{ggggg}
\begin{pmatrix}
\vec{\textbf{p}} & p_0 {\textbf{z}} \\
p_0 {\textbf{z}}^{{\scriptscriptstyle\bigstar}} & -{\textbf{z}}^{{\scriptscriptstyle\bigstar}} \vec{\textbf{p}} {\textbf{z}}
\end{pmatrix}
\equiv \begin{pmatrix} (0,\vec{a}+\vec{j}) & \;\;(d^0,\vec{d}) \\ (d^0,-\vec{d}) & \;\; (0,-\vec{a}+\vec{j}) \end{pmatrix}\,,
\end{eqnarray}
this family of the (co-)adjoint orbits can be characterized through the following conditions:
\begin{eqnarray} \label{conservation laws}
\vec{j} = \frac{1}{d^0} \vec{d}\times\vec{a}\,, \quad 
\kappa^2 = (d^0)^2 + \vec{d}\cdot\vec{d} - \vec{a}\cdot\vec{a} - \vec{j}\cdot\vec{j}\,,
\end{eqnarray}
where the notations `$\cdot$' and `$\times$' denote the Euclidean dot product and the cross product within the context of $\mathbb{R}^{3}$.

Now, given a specific value of $\kappa$, the identification of the corresponding (co-)adjoint orbit as the phase space of a scalar dS$_4$ elementary system allows us to interpret the conditions (\ref{conservation laws}) as the system's conservation laws. To clarify this, we introduce three fundamental quantities: the universal length denoted as $R$, which represents the radius of curvature of the dS$_4$ hyperboloid ${M}_R$; the universal speed of light, symbolized as $c$; and a ``mass" labeled as $m$. In a technical sense, these quantities permit us to assign proper physical dimensions to the variables $(\vec{a},d^0,\vec{d})$ as $\vec{a} = \kappa {\vec{p}}/{mc}$, $d^0 = \kappa {E}/{mc^2}$, $\vec{d} = \kappa {\vec{q}}/{R}$, with $\kappa=mc^2$. Consequently, the conditions (\ref{conservation laws}) can be expressed as follows:
\begin{eqnarray}\label{conservation laws'}
\vec{j} &=& \kappa \; \frac{c}{E R} \vec{l}\,,\quad \mbox{with} \;\;\;\; \vec{l}\equiv\vec{q}\times\vec{p}\,, \nonumber\\
0 &=& E^4 + E^2\left( - m^2c^4 - c^2 (\vec{p}\cdot\vec{p}) + \frac{m^2c^4}{R^2} (\vec{q}\cdot\vec{q}) \right) - \frac{m^2c^6}{R^2} (\vec{l}\cdot\vec{l})\,.
\end{eqnarray}
In the limit of Poincar\'{e} contraction as $R$ approaches infinity, the previously described dS$_4$ construction aligns with the mass shell hyperboloid:
\begin{eqnarray}
E^2 - c^2 (\vec{p}\cdot\vec{p}) = m^2c^4\,,
\end{eqnarray}
which pertains to the description of the co-adjoint orbits for massive scalar elementary systems within the context of Poincaré relativity, as elucidated in reference \cite{Cari1990}.

Note that the phase space for dS$_4$ ``massless" scalar particles can be realized by the ``massless" limit ($\kappa\rightarrow 0$) of the above construction.


\end{document}